# When Is a Crowd Wise?

Clintin P. Davis-Stober, University of Missouri
David V. Budescu, Fordham University
Jason Dana, Yale University
Stephen Broomell, Carnegie Mellon University

1. INTRODUCTION

Numerous studies have established that aggregating judgments or predictions across individuals can be surprisingly accurate in a variety of domains, including prediction markets, political polls, game shows, and forecasting (see Surowiecki, 2004). Under Galton's (1907) conditions of individuals having largely unbiased and independent judgments, the aggregated judgment of a group of individuals is uncontroversially better, on average, than the individual judgments themselves (e.g., Armstrong, 2001; Clemen, 1989; Galton, 1907; Surowiecki, 2004; Winkler, 1971). The boundary conditions of crowd wisdom, however, are not as well-understood. For example, when group members are allowed access to other members' predictions, as opposed to making them independently, their predictions become more positively correlated and the crowd's performance can diminish (Lorenz, Rauhut, Schweitzer, & Helbing, 2011). In the context of handicapping sports results, individuals have been found to make systematically biased predictions, so that their aggregated judgments may not be wise (Simmons, Nelson, Galak, & Frederick, 2011). How robust is crowd wisdom to factors such as non-independence and bias of crowd members' judgments? If the conditions for crowd wisdom are less than ideal, is it better to aggregate judgments or, for instance, rely on a skilled individual judge? Would it be better to add a highly skilled crowd member or a less skilled one who makes systematically different predictions than other members, increasing diversity?

We provide a simple, precise definition of the wisdom-of-the-crowd effect and a systematic way to examine its boundary conditions. We define a crowd as wise if a linear aggregate of its members' judgments of a criterion value has less expected squared error than the judgments of an individual sampled randomly, but not necessarily uniformly, from the crowd. Previous definitions of the wisdom of the crowd effect have largely focused on comparing the crowd's accuracy to that of the average individual member (Larrick, Mannes, & Soll, 2012). Our definition generalizes prior approaches in a couple of ways. We consider crowds created by any linear aggregate, not just simple averaging. Second, our definition allows the comparison of the crowd to an individual selected according to a distribution that could reflect past individual performance, e.g., their skill, or other attributes. On the basis of our definition, we develop a framework for analyzing crowd wisdom that includes various aggregation and sampling rules. These rules include both weighting the aggregate and sampling the individual according to skill, where skill is operationalized as predictive validity, i.e., the correlation between a judge's prediction and the criterion. Although the amount of the crowd's wisdom - the expected difference between individual error and crowd error - is non-linear in the amount of bias and non-independence of the judgments, our results yield simple and general rules specifying when a simple average will be wise. While a simple average of the crowd is not always wise if individuals are not sampled uniformly at random, we show that there always exists some a priori aggregation rule that makes the crowd wise.

## 1.1 Preliminaries

Consider a set of $N$-many decision makers (DMs) where each DM makes a judgment about the unknown value of a criterion. We model the criterion being predicted (or estimated) by the group members as a random variable with finite mean and variance. In this way, we conceptualize our framework as applying to random criteria, as in prediction, as well as to the special cases of estimating a single fixed quantity (which we accommodate by setting the variance of the criterion to 0). We take this criterion value to be a random variable, $\mathcal{Y}$, with mean $\mu_y$ and variance $\sigma_y^2$.

Similarly, we assume that each DM's judgment is a random variable. This assumption represents the variability of a DM who gives variable responses to the same task. With this assumption, we can model how a DM's predictions correlate with the criterion as well as other DMs in the crowd. Let the prediction distribution of the $i^{th}$ DM be the random variable $\mathcal{X}_i$ with mean $\mu_{xi}$ and variance $\sigma_{xi}^2$. A *crowd prediction*, denoted $\mathcal{C}$, is defined as the random variable formed by linearly combining the DMs according to predetermined weights $w_i$, $\mathcal{C} = \sum_{i=1}^{N} w_i \mathcal{X}_i$, with the restriction that all $w_i$'s are non-negative and, to ensure uniqueness, $\sum_{i=1}^{N} w_i = 1$. The weights, $w_i$, are not random variables, but rather fixed choices of how to combine crowd member judgments.

We consider whether the crowd's judgment is expected to be better than an individual crowd member's. Let $\mathcal{P}$ be the random variable formed by selecting a single member of the crowd probabilistically, and let $p_i$ denote the probability of selecting the $i^{th}$ crowd member, with $p_i \geq 0, \forall i \in \{1, 2, \ldots, N\}$ and $\sum_{i=1}^{N} p_i = 1$. As a special case, if all $p_i$ values are equal, i.e., $p_i = \frac{1}{N}, \forall i \in \{1, 2, \ldots, N\}$, then $\mathcal{P}$ reduces to selecting any individual DM with equal probability. Note that we place no a priori restrictions on the $\mu_{xi}$ and $\sigma_{xi}^2$ values. This allows for the possibility that DMs are *biased*, meaning that their average judgment would not equal the average criterion value, $E[\mathcal{X}_i] = \mu_{xi} \neq \mu_y$. Also note that we allow DMs to have different prediction variances, $\sigma_{xi}^2$, and arbitrary covariances with other DMs where $\sigma_{xi,xj}$ denotes the covariance of $\mathcal{X}_i$ and $\mathcal{X}_j$. In other words, the judgments of the crowd members may be correlated with each other.

We define accuracy as the average squared error of prediction (whether a group or individual). This is a common "gold standard" accuracy metric within the field statistics (Lehmann & Casella, 1998). This accuracy metric allows us to derive distribution-free results on crowd wisdom. In other words, we make no assumptions regarding the underlying distributional form of the individual or group's predictions, such as normality, nor do we impose any constraints on the distribution's shape such as symmetry or unimodality. Alternative accuracy definitions (e.g., average absolute error) can change the conclusions of our model, though our approach is one that could, in theory, be extended to any accuracy metric.

## 1.2 Our Model

We define a *wisdom of the crowd effect* to hold if, and only if,

$$E[(\mathcal{C} - \mathcal{Y})^2] \leq E[(\mathcal{P} - \mathcal{Y})^2], \qquad (1)$$

for some crowd aggregate weights, $w_i, i \in \{1, 2, \ldots, N\}$, and selection distribution probability weights $p_i, i \in \{1, 2, \ldots, N\}$. Note that the right-hand side of this inequality is the expected accuracy of selecting an individual according to an arbitrary, pre-specified probability distribution, in contrast to previous formulations such as evaluating the arithmetic mean accuracy of individual predictions (Larrick, Mannes, & Soll, 2012).

The aggregate crowd prediction distribution, $\mathcal{C}$, defined by $w$, has lower expected loss than an individual judgment selected according to the probability measure, $p_i, i \in \{1, 2, \ldots, N\}$, if, and only if, the following inequality holds:

$$(\boldsymbol{\mu}'_X \boldsymbol{w} - \mu_y)^2 + \boldsymbol{w}' \Sigma_{XX} \boldsymbol{w} - 2\boldsymbol{w}' \boldsymbol{\sigma_{xy}} + \sigma_y^2 \leq \sum_{i=1}^{N} p_i [(\mu_{xi} - \mu_y)^2 + \sigma_{xi}^2 - 2\sigma_{yxi} + \sigma_y^2]. \quad (2)$$

This inequality provides an explicit, testable condition to determine whether a given crowd is wise. We demonstrate how this result can be used to evaluate the relative trade-offs of group member interdependence versus bias as well as how to apply this model to data. In the paper, we examine many possible choices of crowd weights and individual selection probabilities.

1.3 Results

Our results suggest that crowd wisdom is robust to different choices of aggregation and sampling rules. That is, how one aggregates the judgments or chooses an individual judge rarely affects the qualitative conclusion that even a crowd that is a simple average of judges is wiser than the individual. By identifying conditions for crowd wisdom, our results also provide guidance for constructing an optimally wise group - a group whose accuracy most exceeds that of its individual members - with two surprising conclusions emerging. First, a crowd becomes wisest when it is maximally informative, which entails that its members' judgments are as *negatively* correlated with each other as possible, as opposed to being independent. Thus, the best judge to add to a crowd is one that is maximally different from others. One intuitive analogy of this result is to think of the group as a financial portfolio: Sometimes it is better to diversify performance by "hedging" and including an asset that performs well when other assets perform poorly. This result provides mathematical support for the idea that crowds with more diversity are wiser (Hong & Page, 2004). Further, our theoretical framework provides a mechanism for determining when it would be better for the overall group prediction to add a group member who, perhaps, is less skilled than the alternative members, but provides diverse predictions. In other words, our framework provides a quantification of the accuracy-diversity trade-off.

A second surprising conclusion is that while the absolute accuracy of the crowd depends on the direction and magnitude of members' bias, it is almost always preferable to use a weighted aggregate of judgments rather than select the single best group member, even if the crowd members are biased. Unless the best group member can be selected deterministically, as in certain intellective tasks (Laughlin, 1996), the decrease in variance of predictions caused by aggregating judgments will offset the bias, a manifestation of the well-known bias/variance tradeoff (Gigone & Hastie, 1997).

We present an application of our framework to experimental studies by re-analyzing the data collected, analyzed and published by Vul and Pashler (2008) and Simmons, Nelson, Galak, and Frederick (2011). Our analysis finds a "wisdom of the crowd" effect when applied to the group of individuals from Vul and Pashler (2008), extending the original analysis which examined the accuracy of pooled repeated judgments *within* individuals. Our re-analysis of the Simmons et al. (2011) data supports the overall treatment effect of increasing individual bias by manipulating the sports betting information available to them. In contrast to the original findings reported by Simmons et al. (2011), our re-analysis, guided by our new formulation, finds an overall improvement of the crowd's predictions relative to individuals across all treatments in the study. In other words, while the members are individually biased and the crowd not particularly accurate, the crowd is still wise relative to the individual.